\documentclass[aps, prl, reprint, longbibliography, nofootinbib, floatfix]{revtex4-2}
\pdfoutput=1
\usepackage{amsmath,amssymb,amsfonts}
\usepackage{mathrsfs}
\usepackage{bbm}
\usepackage{slashed}
\usepackage{graphicx}
\usepackage{verbatim}
\usepackage[T1]{fontenc}
\usepackage[utf8]{inputenc}
\usepackage[colorlinks]{hyperref}
\usepackage[english]{babel}
\usepackage{mathbbol}
\usepackage[dvipsnames]{xcolor}
\usepackage[normalem]{ulem}
\usepackage{svg}
\usepackage{graphicx} 

\hypersetup{
	colorlinks,
	linkcolor={blue!75!black},
	citecolor={blue!75!black},
	urlcolor={blue!75!black}
}

\newcommand{\GR}{{\small GR}}
\newcommand{\eg}{{\textit{e.g.}}}
\newcommand{\ie}{{\textit{i.e.}}}

\newcommand{\GN}{\ensuremath{G_N}}
\newcommand{\mass}{\ensuremath{M}}

\allowdisplaybreaks

\begin{document}
\title{Sifting quantum black holes through the principle of least action}
\author{Benjamin Knorr\,\href{https://orcid.org/0000-0001-6700-6501}{\protect  \includegraphics[scale=.045]{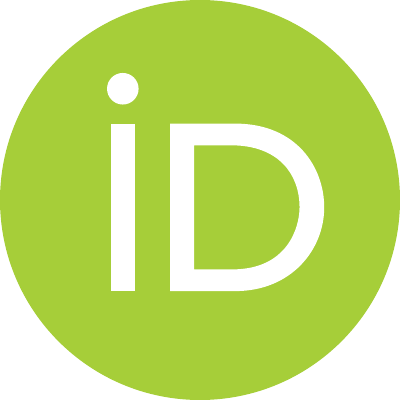}}}
\email[]{bknorr@perimeterinstitute.ca}
\author{Alessia Platania\,\href{https://orcid.org/0000-0001-7789-344X}{\protect  \includegraphics[scale=.045]{ORCIDiD_iconvector}}}
\email[]{aplatania@perimeterinstitute.ca}
\affiliation{
Perimeter Institute for Theoretical Physics, 31 Caroline Street North, Waterloo, ON N2L 2Y5, Canada
}

\begin{abstract}
 We tackle the question of whether regular black holes or other alternatives to the Schwarzschild solution can arise from an action principle in quantum gravity. Focusing on an asymptotic expansion of such solutions and inspecting the corresponding field equations, we demonstrate that their realization within a principle of stationary action would require either fine-tuning, or strong infrared non-localities in the gravitational effective action. This points to an incompatibility between large-distance locality and many of the proposed alternatives to classical black holes.
\end{abstract}

\maketitle

\textit{Introduction.} --- Black holes are some of the most fascinating objects in our Universe. Our current understanding of them is based on the Schwarzschild solution of General Relativity (\GR{}), and its Kerr and Reissner-Nordstr\"om generalizations. Since the discovery of these solutions, and latest with the derivation of singularity theorems~\cite{Penrose:1964wq, Hawking:1969sw}, there has been a heated discussion about the singularities of classical black holes. The general expectation is that quantum gravity should ultimately resolve any singularities. Yet, while the inside of black holes is still under theoretical investigation, their macroscopic properties and shape are currently in the spotlight of astrophysical experiments and observations~\cite{LIGOScientific:2016sjg, EventHorizonTelescope:2019dse}.

The construction of a well-defined, predictive and falsifiable theory of quantum gravity has still not succeeded. Nonetheless, regular alternatives to classical black holes have been advanced, both by investigating simplified quantum gravitational settings, as well as within model-building approaches applying a ``singularity-resolution principle''. For instance, in some realizations of string theory, classical black holes could be replaced by fuzzballs~\cite{Mathur:2005zp, Chen:2014loa, Bena:2016ypk} or by wormholes~\cite{Maldacena:2017axo, Marolf:2021kjc, Guo:2021blh}. Compact objects such as quasi-black holes~\cite{Lemos:2007yh, Barcelo:2007yk} and gravastars~\cite{Mazur:2001fv} have also been proposed as viable alternatives to black holes. Other theories of quantum gravity, including loop quantum gravity~\cite{Ashtekar:2005qt, Gambini:2013ooa, Rovelli:2014cta, DeLorenzo:2014pta, Perez:2017cmj, Bojowald:2020dkb} and asymptotically safe gravity~\cite{Bonanno:2000ep, Koch:2013owa, Pawlowski:2018swz, Platania:2019kyx, Bosma:2019aiu, Borissova:2020knn} seem instead to point to ``regularized'' versions of classical black holes, with modifications occurring at Planckian ``distances'' from the would-be singularity. Well-known examples of regular black hole models include the Dymnikova~\cite{Dymnikova:1992ux}, Bardeen~\cite{Bardeen:1968}, Bonanno-Reuter~\cite{Bonanno:2000ep}, and Hayward~\cite{Hayward:2005gi} black holes.\footnote{The viability of these specific models is still uncertain due to potential instabilities of the inner horizon, and currently under debate~\cite{Carballo-Rubio:2018pmi, Bonanno:2020fgp, Carballo-Rubio:2021bpr}. Nonetheless, it is conceivable that the dynamics of a gravitational collapse could have an important impact on this question, since it could result in a black hole spacetime with integrable singularities~\cite{Strokov:2016wco, Bonanno:2016dyv, Rignon-Bret:2021jch} or in a compact object~\cite{Barcelo:2007yk}, rather than in a regular black hole. At any rate, the analysis in the present Letter is independent of the outcome of the aforementioned debate.} Regardless of the specific quantum gravity theory, it is a key question whether spacetimes deviating from the classical Schwarzschild metric can be found as solutions to effective field equations stemming from an action principle in quantum gravity.\footnote{It is known that certain regular black holes can be found from an action principle coupling \GR{} to non-linear electrodynamics~\cite{Dymnikova:2004zc} or other exotic modifications involving magnetic monopoles~\cite{Ayon-Beato:2000mjt}. However, such modifications have not found experimental confirmation so far. It thus seems unlikely that singularity resolution in General Relativity would arise from deviations to Maxwell theory, rather than from quantum gravity.} This is the topic of this Letter.

In the following, we assume that all phenomena in the Universe, including all matter and interactions, can be described within a quantum field theoretic framework in terms of an effective action. We then imagine to integrate out all matter and gauge fields, as well as quantum-gravitational fluctuations, so that the resulting effective action is a functional of the metric only. The question that we attempt to address is whether one can find an effective action whose equations of motion are solved by one of the known alternatives to classical black holes. Concretely, we will focus on the large-distance behavior of static, spherically symmetric spacetimes, whose metric coefficients admit an asymptotic expansion in powers of the radial coordinate, and we will attempt to determine approximations to the corresponding gravitational effective action. Such asymptotic corrections to the Schwarzschild metric are expected, since quantum-gravitational fluctuations modify the effective field equations by higher-order curvature operators in such a way that Ricci-flat spacetimes are no longer solutions. A prime example for such a higher-derivative correction is the Goroff-Sagnotti term~\cite{Goroff:1985sz,Goroff:1985th}. We remark in passing that, even though regular black holes constitute the primary motivation of our work, our analysis applies to any spherically symmetric, asymptotically flat spacetime, regular or not, with or without event horizons.

Our result is that power-law corrections to the large distance behavior of the Schwarzschild metric can be generated by local higher derivative terms, if the corresponding leading-order exponent satisfies a certain lower bound. For all other power-law corrections, one generically needs specific non-local terms of the Polyakov type~\cite{Polyakov:1981rd} in the effective action. Surprisingly, in this case the leading-order correction removes the entire kinetic term of the graviton in Minkowski space, independent of the exponent of the correction. Alternatively, one would need even stronger infrared non-localities at higher orders in the curvature expansion. As a key consequence, most of the known alternatives to classical black holes would likely require infrared modifications of \GR{}.

\textit{Setup.} --- Regular black holes can be interpreted as solutions to modified Einstein field equations,
\begin{equation}
    G_{\mu\nu}=8\pi \GN{} T_{\mu\nu}^\mathrm{eff} \,,
\end{equation}
with~$T_{\mu\nu}^\mathrm{eff}$ being an effective energy-momentum tensor, encoding quantum gravitational and matter effects or exotic ``new physics''. An open, intriguing question is whether these effective field equations can arise from a principle of least action. If singularity resolution is to be attributed to quantum gravity, regardless of the specific model, this action ought to be a gravitational effective action~$\Gamma[g]$. The corresponding regular metric $g$ would thus be a solution to the corresponding effective field equations.

On these grounds, our investigation is based on an effective action~$\Gamma$, which includes all quantum effects. The quantum equation of motion then reads
\begin{equation}\label{eq:qEoM}
    \frac{\delta\Gamma[g]}{\delta g_{\mu\nu}} = 0 \, .
\end{equation}
We will investigate a perturbed Schwarzschild metric in Schwarzschild coordinates,\footnote{This is the most general spherically symmetric, asymptotically flat spacetime in spherical coordinates. Even modifying the angular metric coefficients by a multiplicative function $C(\tilde{r})$, as in~\cite{Simpson:2018tsi} and in most wormholes spacetimes, one can always perform a coordinate transformation $\tilde{r}\to r=\tilde{r}C(\tilde{r})$ to recast the metric into the form~\eqref{eq:metric}.}
\begin{equation}\label{eq:metric}
    g_{\mu\nu} = \text{diag} \left( -f_{tt}(r), \frac{1}{f_{rr}(r)}, r^2, r^2 \sin\theta \right) \, ,
\end{equation}
with
\begin{equation}\label{eq:metricfunctions}
    f_{tt}(r) \sim 1 - \frac{2\GN{} \mass{}}{r} + \frac{c_t}{r^{n_t}} \, , \,\, f_{rr}(r) \sim 1 - \frac{2\GN{} \mass{}}{r} + \frac{c_r}{r^{n_r}} \, .
\end{equation}
Here, \GN{} is Newton's constant and \mass{} is the mass of the corresponding Schwarzschild black hole. Our ansatz for the perturbed Schwarzschild metric thus only applies to modified metrics admitting such an asymptotic expansion, \eg, the Bardeen~\cite{Bardeen:1968}, Bonanno-Reuter~\cite{Bonanno:2000ep}, Hayward~\cite{Hayward:2005gi} and Simpson-Visser~\cite{Simpson:2018tsi} spacetimes. All asymptotic relations in this work are understood in the limit~$r\to\infty$, which is the focus of our investigation. The correction terms are assumed to be sub-leading in this limit so that~$n_r, n_t > 1$, but otherwise we put no constraints on the power laws. In this regime, a curvature expansion of the effective action is expected to be valid, as long as strong non-localities are absent. This entails that our ansatz for~$\Gamma$ reads\footnote{A similar term quadratic in the Weyl tensor can be absorbed in the two present quadratic terms, cubic terms, and the topological Euler characteristic via the Bianchi identity~\cite{Avramidi:2000bm}.}
\begin{equation}\label{eq:EAansatz}
\begin{aligned}
    \Gamma &= \frac{1}{16\pi \GN{}} \int \text{d}^4x \, \sqrt{-g} \bigg[ -R - \frac{1}{6} R f_R(\Delta) R 
    \\
    &\hspace{2.5cm} + R^\mu_{\phantom{\mu}\nu}f_{Ric}(\Delta)R^\nu_{\phantom{\nu}\mu} + \mathcal O(\mathcal R^3) \bigg] \, ,
\end{aligned}
\end{equation}
where~$\Delta=-g^{\mu\nu}D_\mu D_\nu$ is the d'Alembert operator of the metric~$g$, and~$\mathcal R$ denotes a generic curvature tensor. The operator functions~$f_R$ and~$f_{Ric}$ are known as form factors, and are related to the physical renormalization group running, \ie{} the momentum dependence of the graviton propagator in Minkowski space. Efforts to compute them from first principles can be found in~\cite{Christiansen:2014raa, Knorr:2018kog, Bosma:2019aiu, Knorr:2019atm, Knorr:2021niv, Bonanno:2021squ, Fehre:2021eob}. Our aim is to constrain the form factors by assuming that the metric~\eqref{eq:metric} with coefficients~\eqref{eq:metricfunctions} solves the effective field equations~\eqref{eq:qEoM} originating from \eqref{eq:EAansatz} for large~$r$.

To simplify the presentation, in the following we will assume that~$n_r = n_t \equiv n$, so that the radial and the temporal component of the metric show the same sub-leading behavior. We performed the full calculation with unequal power laws and arrive at the same leading order result, with~$n=\min(n_r,n_t)$.

Since in the action principle $\delta\Gamma=0$, the variation acts linearly on the effective action, $\delta \Gamma=\delta \Gamma_{GR}+\delta \Gamma_{\mathcal{R}^2}+\delta \Gamma_{\mathcal{R}^3}+\dots$, we will discuss the contribution of the individual terms in the action to the equations of motion independently. 

\textit{\GR{} equations of motion.} --- The first step to derive constraints for the form factors is to compute the contribution $\delta \Gamma_{GR}$ of the \GR{} action to the equations of motion. We can focus on the~$(rr)$ and~$(tt)$ components, since the other components either vanish or are dependent. For large~$r$, we find to leading order,
\begin{equation}\label{eq:GReom}
    \left. \left\{ \frac{\delta\Gamma}{\delta g_{tt}}, \frac{\delta\Gamma}{\delta g_{rr}} \right\} \right|_{GR} \sim \, \frac{r^{-n-2}}{16\pi \GN{}} \left\{ c_r(n-1), (c_r-n \, c_t) \right\} \, .
\end{equation}
These terms have to be canceled by those generated by higher order terms in the effective action.

\textit{Local corrections.} --- The simplest choice for the form factors is a truncated Taylor expansion.\footnote{Such an artificial truncation could lead to the appearance of fictitious ghosts, but this does not necessarily break unitarity in the full quantum theory~\cite{Platania:2020knd}.} This corresponds to a local modification to the Einstein-Hilbert action. To lowest order, the form factors are then constants and simply correspond to the Stelle terms~\cite{Stelle:1976gc, Stelle:1977ry}, $\delta \Gamma_{\mathcal{R}^2}^{\text{loc}}\simeq \delta \Gamma_\mathrm{Stelle}$. These terms however yield a different leading-order asymptotic power law,
\begin{equation}\label{eq:StelleEoM}
    \left.  \frac{\delta\Gamma^{\text{loc}}}{\delta g_{tt,rr}} \right|_{\text{Stelle}} \sim a_{tt,rr} \, r^{-n-4} \, ,
\end{equation}
for some computable constants~$a_{tt,rr}$. Including terms with positive powers of the d'Alembert operator yields contributions which are even more sub-leading. We thus conclude that these local terms cannot cancel the \GR{} terms~\eqref{eq:GReom}.

Similar arguments also apply to \emph{all} local terms with three or more powers of the curvature. This is because we consider an expansion of an asymptotically flat spacetime about infinite radial distance where all curvature tensors vanish. This implies that the more curvature tensors an expression has, the more sub-leading its contribution to the field equations is in this limit. Moreover, since the first contribution of the Ricci tensor to the asymptotic expansion contains a factor of~$n$ in the exponent of the power law in~$1/r$, terms with more than one Ricci tensor or scalar can never cancel the \GR{} contribution for~$n>0$.

The only terms leading to an $n$-independent power law are those constructed from the Weyl tensor and at most one occurrence of the Ricci tensor or scalar: the Weyl tensor of the Schwarzschild-part of the metric is non-zero, and hence the leading-order contribution to the field equations is an~$n$-independent power law. Such terms can thus cancel the \GR{} contribution~\eqref{eq:GReom} for specific values of~$n$.\footnote{The~$C^2$ term is an exception in~$d=4$, as it can be rewritten in terms of the topological Euler term and a combination of squared Ricci scalars and tensors.} The lowest order correction for which such a cancellation is possible is
\begin{equation}
\begin{aligned}
    \Gamma_{\mathcal R^3}^{\text{loc}} = \frac{1}{16\pi \GN{}} \int \text{d}^4x \,&  \sqrt{-g} \, \bigg[ k_{RC^2} \, R \, C^{\mu\nu\rho\sigma} C_{\mu\nu\rho\sigma} \\
    &+ k_{C^3} \, C_{\mu\nu}^{\phantom{\mu\nu}\rho\sigma} C_{\rho\sigma}^{\phantom{\rho\sigma}\tau\omega} C_{\tau\omega}^{\phantom{\tau\omega}\mu\nu} \bigg] \, .
\end{aligned}
\end{equation}
Its leading order contribution to the equations of motion has the form
\begin{equation}\label{eq:r-8part}
    \left. \frac{\delta\Gamma^{\text{loc}}}{\delta g_{tt,rr}} \right|_{\mathcal R^3} \sim b_{tt,rr} \, r^{-8} \, ,
\end{equation}
where~$b_{tt,rr}$ are constants which depend on the cubic couplings~$k_{RC^2}, k_{C^3}$. This combination of terms thereby only allows a cancellation for the case~$n=6$, with the specific choice
\begin{equation}
    k_{RC^2} = - \frac{c_r - 3c_t}{432 \GN{}^2 \mass{}^2} \, , \qquad k_{C^3} = - \frac{c_r + 6c_t}{216 \GN{}^2 \mass{}^2} \, .
\end{equation}
This case has also been investigated in more detail in~\cite{Anselmi:2013wha, deRham:2020ejn}. It is clear that even higher order terms allow for cancellations for other specific powers~$n$. In fact, a cancellation is possible for all integers~$n$ with~$n\geq 6$,~$n\neq 7$.\footnote{Let us remark that the gravitational potential defined from the geodesic equation, which is related to the $g_{tt}$-component of the metric, and the one obtained from a scattering amplitude~\cite{Bjerrum-Bohr:2002gqz}, need not to agree in general. Indeed, while the former is the potential acting on an idealized, non-interacting test particle, the latter is the potential on an interacting scalar field, and thus accounts for quantum corrections to gravity-matter vertices which by construction are not included in $g_{tt}$. As an important example, the potential derived from the scattering of two massive scalars on a de Sitter background~\cite{Ferrero:2021lhd} is manifestly different from the $g_{tt}$-component of the Schwarzschild-de Sitter spacetime. Therefore, our constraints do not apply to the leading-order corrections to the Newtonian potential found in~\cite{Donoghue:1993eb}.} This can be seen by considering monomials of the form~$\left[(\Delta^m R) (C^2)^k (C^3)^l\right]$ for integers~$k,l,m$, or similar monomials with different contractions and distributions of covariant derivatives, as they contribute to the field equations with a leading-order power law of the form~$\sim r^{-2 - 6 k - 9 l - 2 m}$. Thus, upon confining to local operators only, our analysis sets an important lower bound on the value of~$n$. As a consequence, well-known black hole models as the Hayward spacetime, with
\begin{equation}
\begin{aligned}
    f_{tt}(r)=f_{rr}(r) & = 1-\frac{2 \mass{} \GN{}}{r}\frac{r^3}{r^3+2 \mass{} \GN{}^{2}} \\
    &\sim 1-\frac{2 \mass{} \GN{}}{r}+\frac{4\GN{}^3 \mass{}^2}{r^4}\,,
\end{aligned}
\end{equation}
cannot be solutions to local gravitational effective field equations. A Hayward-like solution potentially compatible with a stationary local-action principle would rather be
\begin{equation}
    f_{tt}(r)=f_{rr}(r) = 1-\frac{2 \mass{} \GN{}}{r}\frac{r^5}{r^5+2 \mass{} \GN{}^{3}}\,.
\end{equation}
Similar considerations apply to other alternatives to classical black holes that admit an asymptotic expansion of the form~\eqref{eq:metricfunctions}, e.g., those discussed in~\cite{Bardeen:1968, Bonanno:2000ep, Hayward:2005gi}.

Let us remark that, regardless of the specific power~$n$, in order for the metric~\eqref{eq:metricfunctions} to be a solution to the full field equations, other higher-derivative operators ought to yield terms in the field equations which cancel both the next-to-leading contributions to the scaling~$\sim r^{-2 - 6 k - 9 l - 2 m}$, as well as the leading-order scaling produced by the lower-derivative operators (as those in Eq.~\eqref{eq:StelleEoM} and Eq.~\eqref{eq:r-8part}) eventually appearing in the effective action. Thus, requiring regular spacetimes compatible with the asymptotic expansion~\eqref{eq:metricfunctions} to arise from a principle of least (local) action translates into strong constraints on both the order $n$ of the corrections and the values of the couplings of the local operators appearing in the effective action. As a final remark, we note that even when modifying our ansatz for the metric coefficients~\eqref{eq:metricfunctions} by including the Yukawa terms in~\cite{Lu:2015psa}, the constraints we have derived remain valid. Indeed, the Yukawa terms in the metric would contribute to the asymptotic expansion of the field equation with additional exponential terms whose overall coefficient must vanish independently of those of simple power laws. Consequently, this would lead to additional, independent constraints, and would not modify those we have derived here.

A cancellation for generic $n$ and the avoidance of the aforementioned fine-tuning require the presence of non-local terms in the effective action, which we will discuss next.

\textit{Non-local corrections} --- The scaling of the correction terms in Eq.~\eqref{eq:StelleEoM} stemming from Stelle gravity suggest that canceling the \GR{} terms in Eq.~\eqref{eq:GReom} can be achieved by form factors in $\delta \Gamma_{\mathcal{R}^2}$ with an inverse power of the argument. Specifically, let us consider
\begin{equation}\label{eq:nonlocalFF}
    f_R(\Delta) = \frac{\alpha}{\Delta} \, , \qquad f_{Ric}(\Delta) = \frac{\beta}{\Delta} \, .
\end{equation}
Such terms have been studied before in the context of cosmology \cite{Wetterich:1997bz, Deser:2007jk, Nersisyan:2016jta}.
By naive counting of derivatives, this choice has the chance of providing a contribution to the equations of motion which can cancel~\eqref{eq:GReom} for arbitrary~$n$. Notably, these terms are not of the form expected from effective field theory~\cite{Donoghue:1993eb}, which would rather come in the form of logarithms. We have checked that logarithms do not give rise to the correct power law contribution to the equations of motion.

For both the Ricci scalar and the Ricci tensor term, the leading-order contribution to the equations of motion comes from the variation of one of the curvature tensors. All other variations are easily seen to come with sub-leading power laws, which we also have verified explicitly. As a consequence, their leading-order contributions to the equations of motion are structurally of the form $D D \Delta^{-1} R$. Thus, in order to determine the contribution to the field equations stemming from the non-local, quadratic part of the action, the next step is to evaluate the action of the inverse d'Alembertian operator on both the Ricci scalar and the Ricci tensor.

Non-local operators such as $\Delta^{-1}$ bring several technical difficulties with them. For instance, in order to define them, one has to specify boundary conditions. In our case these are naturally tied to the zero modes of the corresponding operators. Specifically, we impose that in the inversion of the operator, no zero mode contributions are added. The reason behind this choice is that in the limit of a flat spacetime where the curvature terms vanish, the zero modes would attribute a finite value to acting with an inverse d'Alembertian on them. 

A second difficulty is to actually compute the action of the inverse d'Alembertian operator on a curvature tensor. In the case of $f_R(\Delta)$, we define $\Delta^{-1}R$ by solving the equation $\Delta\chi = R$ for $\chi$, with the aforementioned boundary conditions.
This equation can be solved in closed form for arbitrary static, spherically symmetric metrics. With our boundary conditions imposed, the solution reads
\begin{equation}
    \chi(r) = \int_r^\infty \text{d}x \frac{-1}{x^2 \sqrt{f_{rr}(x) f_{tt}(x)}} \int_x^\infty \text{d}y \sqrt{\frac{f_{tt}(y)}{f_{rr}(y)}} y^2 R(y) \, ,
\end{equation}
where $R(y)$ is the expression of the Ricci scalar for the metric~\eqref{eq:metric} as a function of the radial coordinate $y\equiv r$. The integrals converge as long as $r$ is large enough, since $R(r)\sim (n-1)(2c_r-n c_t)r^{-n-2}$ for our ansatz~\eqref{eq:metricfunctions} and $n>1$ by assumption. From this expression, it is straightforward to derive an asymptotic expansion for $\chi$.

A similar procedure can be carried out for the case of the Ricci tensor. In this case the inversion of the d'Alembertian requires solving $\Delta \Sigma^\mu_{\phantom{\mu}\nu} = R^\mu_{\phantom{\mu}\nu}$ for the tensor $\Sigma$. The index structure was chosen to simplify the calculation. Indeed, the symmetries of~$R^\mu_{\phantom{\mu}\nu}$ and the spherical symmetry of the metric~\eqref{eq:metric} imply that~$\Sigma^\mu_{\phantom{\mu}\nu}$ must be diagonal, with~$\Sigma^\theta_{\phantom{\theta}\theta}=\Sigma^\varphi_{\phantom{\varphi}\varphi}$. Unfortunately, we were not able to solve this equation analytically, and thus we resorted to a direct asymptotic expansion to compute~$\Sigma$.

We refrain from presenting the full details of the calculation since not much can be learned from them. As an intermediate result, we find that for large~$r$,
\begin{align}
    \frac{1}{\Delta} R &\sim -\frac{r^2}{n(n-1)} R \sim \frac{c_t - \frac{2}{n} c_r}{r^n} \, , \\
    \frac{1}{\Delta} R^\mu_{\phantom{\mu}\nu} & \sim \, r^{-n} \mathrm{diag}\left( \frac{c_t}{2}, \frac{c_t n (n-1) - 2 c_r (n-2)}{2n (n-3)},\right. \nonumber \\ 
    & \hspace{1cm} \left. -\frac{c_r (n-4) + c_t n}{2 n (n-3)}, - \frac{c_r (n-4) + c_t n}{2 n (n-3)} \right) \,.
\end{align}
Matching the resulting contribution to the equations of motion to cancel the terms in Eq.~\eqref{eq:GReom}, we find
\begin{equation}\label{eq:nonlocalcoeffsol}
    \alpha = -3 \, , \qquad \beta = -1 \, ,
\end{equation}
Surprisingly, these effective couplings are independent of~$n$. Let us note that the appearance of the factor $(n-3)$ (and $(n-2)$ in sub-leading orders) in the denominators of $\Sigma$ simply indicate that for $n=2,3$, logarithmic terms have to be included in the large-$r$ expansion of $\Sigma$. The values of $\alpha$ and $\beta$ do not change in these cases.

This result has astonishing consequences for the spectrum of the theory, since it implies a vanishing graviton two-point function in Minkowski space, and thereby a diverging graviton propagator. In other words, the graviton would not propagate in this theory.\footnote{Note that introducing higher-order derivative terms in the action could in principle make the graviton two-point function non-vanishing. Nonetheless, this would not be sufficient to restore the standard leading-order dispersion relation for the graviton.} This makes the class of metrics~\eqref{eq:metric} with coefficients~\eqref{eq:metricfunctions} physically unacceptable: asymptotically flat black hole spacetimes admitting the asymptotic expansion~\eqref{eq:metricfunctions} would require either fine-tuning of both the power-law exponent and the coefficients of the effective action, or a theory characterized by large-distance non-localities of the form~\eqref{eq:nonlocalFF} and by a non-propagating graviton.

Alternatively, in order to cancel the \GR{} terms \eqref{eq:GReom} while keeping the standard graviton dispersion relation, one could trade the Polyakov terms~\eqref{eq:nonlocalFF} for stronger infrared non-localities at higher orders in the curvature expansion~\eqref{eq:EAansatz}, \eg{} of the form $\mathcal{R}^2 \Delta^{-\frac{n}{2}-2} \mathcal R$. However, such non-localities would likely produce observable deviations from \GR{} at large distance scales.

Our results thus indicate that, as long as the cosmological constant is negligible, spacetimes with~$1<n<6$, including known (spherically symmetric) regular black holes and wormholes~\cite{Bardeen:1968, Hayward:2005gi, Simpson:2018tsi} are ruled out by our considerations.

\textit{Conclusions.} --- Our Letter highlights novel, non-trivial restrictions on the class of modifications to the Schwarzschild geometry compatible with large-distance locality and a principle of stationary action in quantum gravity.

Enforcing the validity of a stationary action principle poses restrictions on the leading-order correction to Schwarzschild black holes at large distances. Spacetimes with algebraic metric components which do not satisfy these constraints cannot be realized within a static, spherically symmetric setup. Their realization within a principle of least action would entail the existence of large-distance non-localities of the Polyakov type, as well as a vanishing graviton two-point function in Minkowski spacetimes -- which in turn would forbid the standard propagation of gravitational waves -- or even stronger infrared non-localities at higher order in a curvature expansion of the effective action. Assuming the validity of a principle of least action for gravity, the latter class of spacetimes appears to be ruled out. This includes the well-known Hayward and Bardeen black holes.

Should our results extend to the rotating case, they would point at one of the following possibilities:
\begin{itemize}
 \item black holes have algebraic metric coefficients whose asymptotic expansion obeys a specific power-law behavior~\cite{Anselmi:2013wha, deRham:2020ejn},
 \item black holes are realized in the form of Dymnikova spacetimes, or analogous black holes or wormholes with transcendental metric components, as it happens \eg{} for black holes in quadratic gravity~\cite{Lu:2015cqa, Lu:2015tle, Lu:2015psa},
 \item the cosmological constant, albeit tiny, restores compatibility of regular black holes with algebraic metric components and locality at large distance for arbitrary $n$, or
 \item the metric description of gravity, quantum field theory, or the principle of least action fail to provide an accurate description of our universe, even at large distances. %(?italianemoji?).
\end{itemize}

Our results resonate with the conclusions of~\cite{Maeda:2021jdc}, as well as with novel expectations that even Planck-scale modifications of classical black holes could have an impact on large-scale physics~\cite{Carballo-Rubio:2019nel}. Our investigation thus sheds new light on the realization of alternatives to classical black holes within metric approaches to quantum gravity.

\bigskip

\acknowledgments

The authors would like to thank N.~Afshordi, I.~Basile, A.~Bonanno, L.~Buoninfante, R.~Casadio, A.~Eichhorn and K.~Stelle for interesting discussions. The authors acknowledge support by Perimeter Institute for Theoretical Physics. Research at Perimeter Institute is supported in part by the Government of Canada through the Department of Innovation, Science and Economic Development and by the Province of Ontario through the Ministry of Colleges and Universities.

\bibliographystyle{apsrev4-2}
\bibliography{general_bib}

\end{document}